\documentclass[prl,twocolumn,showpacs,amsmath,amssymb,superscriptaddress]{revtex4}
\usepackage{graphicx}
\begin{document}

\title{Reply to comment on "Exact Solution for Three-Dimensional Ising Model" added by Reply to the Reply by Jacques H.H. Perk}
\author{Degang Zhang}
\affiliation{College of Physics and Electronic Engineering, Sichuan Normal University,
Chengdu 610101, China}
\affiliation{Institute of Solid State Physics, Sichuan Normal
University, Chengdu 610101, China}

\begin{abstract}

In this reply, I point out that the comment by Jacques H.H. Perk on my paper "Exact Solution for Three-Dimensional Ising Model" [Symmetry 2021, 13, 1837] is wrong because the author seems not to understand the Onsager operators along both {\bf X} and {\bf Y} directions and the relation between them.

\end{abstract}

\maketitle

It is known that the three-dimensional (3D) classical Ising model can be exactly solved by employing the transfer matrix method. When this approach is applied, the Hamiltonian of 3D classical Ising model becomes the 2D transverse quantum Ising model on the transfer matrix plane. For the spin configurations on a plane, there are two kinds of the boundary conditions: (I) the periodic boundary conditions along both {\bf X} and {\bf Y} directions, and (II) the screw boundary condition along {\bf X} direction and the periodic boundary condition along {\bf Y} direction. Obviously, the Hamiltonian of the 2D quantum Ising model with (II) is composed of that with (I) and an alternating Ising chain with the periodic boundary condition along {\bf Y} direction. I note that such a classical Ising chain could not affect the  thermodynamic quantities of the 3D Ising model in the bulk. However, it is nothing but the screw boundary condition along {\bf X} direction that makes the local Lie algebra produced by the Onsager operators closed.

In Ref. [1], I have solved exactly the 3D Ising model with (II) by an algebraic approach, similar to that in solving the 2D Ising model [2]. The analytical expression of the partition function per atom for the infinite crystal is presented. The critical temperatures associated with three magnetization directions are determined. Some  thermodynamic quantities, e.g. the free energy, the internal energy, and the specific heat, and  the high-temperature expansions of the partition function per atom with $J_1=J_2=J$, are also calculated. In addition, the critical exponents $\alpha_{3D}=0, \nu_{3D}=\frac{2}{3}$ and $\mu_{3D}=\frac{4}{3}$ are obtained. Here I emphasize that for (I), the Onsager operators along X or Y direction do not construct a Lie algebra, and hence the 3D Ising model with (I) cannot be solved exactly. Although the boundary conditions play a crucial role in solving 3D Ising model, the physical properties are irrelevant to them in the thermodynamic limit.

In the comment [3], Jacques H.H. Perk does not understand the operators ${\cal L}^p_{a,a+1}$, $L^p_{a,a+1}$, ${\cal A}_{p,1}$ and $A_{p,1}$ and the relation between them. Here $p$ ($p=1, 2, \cdots, m)$ denotes the $p$th subspace with $2^{2n}$ dimensions along {\bf Y} direction in Fig. 1 [1]. ${\cal A}_{p,1}\equiv A_{p,1}$ comes from the fact that $[{\cal A}_{p,1},A_{p,1}]=0$ and both ${\cal A}_{p,1}$ and $A_{p,1}$ satisfy the same local Lie algebra [1] and can be easily proved by defining new spin variables.
Due to ${\cal A}_{p,1}=\sum_{a=1}^{n}{\cal L}^p_{a,a+1}$ and $A_{p,1}=\sum_{a=1}^{n}L^p_{a,a+1}$, where ${\cal L}^p_{a,a+1}$ in $2^{2n}$ subspace and $L^p_{a,a+1}$ in $2^{2mn}$ space are the Onsager operators along {\bf Y} and {\bf X} directions, respectively (see Fig. 2 in Ref. [1]), then
$A_{p,1}=\sum_{a=1}^{n}L^p_{a,a+1}=\sum_{a=1}^{n}L_{p+(a-1)m,p+am}
=\sum_{a=1}^{n}\sigma^z_{p+(a-1)m}(\prod_{i=p+(a-1)m+1}^{p+am-1}\sigma^x_i)\sigma^z_{p+am}
=\sum_{a=1}^{n}\sigma^z_{p+(a-1)m}{\sigma^\prime}^z_{p+am}=\sum_{a=1}^{n}{\cal L}^p_{a,a+1}={\cal A}_{p,1}$,
where ${\sigma^\prime}^z_{p+am}
=(\prod_{i=p+(a-1)m+1}^{p+am-1}\sigma^x_i)\sigma^z_{p+am}$. Because $[\prod_{i=p+(a-1)m+1}^{p+am-1}\sigma^x_i,
{\cal A}_{p,1}]=[\prod_{i=p+(a-1)m+1}^{p+am-1}\sigma^x_i,A_{p,1}]=0$,
$\prod_{i=p+(a-1)m+1}^{p+am-1}\sigma^x_i$ are conserved quanities.
Compared to the equation above, Eq. (2) in Ref. [3] is obviously wrong! In addition, the operators $\sigma^z_p\sigma^z_{p+m}$ and $\sigma^z_p\sigma^x_{p+1}\cdots\sigma^x_{p+m-1}\sigma^z_{p+m}$ in Eq. (2) in Ref. [3] are not used in my paper [1]. Furthermore, because the eigenvalue of $\sigma^x_{p+1}\cdots\sigma^x_{p+m-1}$ is 1 or $-1$, $\sigma^z_p\sigma^z_{p+m}\not=\sigma^z_p\sigma^x_{p+1}\cdots\sigma^x_{p+m-1}\sigma^z_{p+m}$. Therefore, such a fatal error in Eq. (2) in Ref. [3] should lead to the ridiculous results.

It is noted that the exact partition function of 3D Ising model in the thermodynamic limit oscillates with the size $m$ along {\bf X} direction [1]. However, all the previous numerical calculations for the physical quantities, including the critical temperatures, did not take the oscillatory effect into account [4]. Therefore, the conjectured values extrapolating to the infinite system in the numerical calculations seem to be inaccurate, and the 3D finite-size scaling theory must be modified. Here I emphasize
that the exact critical temperatures at the ferromagnetic and paramagnetic phase transitions determined by Eq. (32) in Ref. [1] coincide completely with the results found in the domain wall analysis [5] and the asymptotically exact values in the anisotropic limit [6,7].

The specific heat $C$ of 2D Ising model with the quadratic symmetry (i.e. $H_1=H_{2D}$) have been calculated
analytically by Onsager and can be expressed in terms of the complete elliptic integrals [2]. The critical exponent associated
with the specific heat $\alpha_{2D}=0$ due to the log divergence at the critical point [8].
Because 3D Ising model with the simple cubic symmetry (i.e. $H_1=H_2=H$) can be mapped exactly
into 2D one by Eq. (41) in Ref. [1], the expression of $C$ in three dimensions has similar form with that in two dimensions. So the critical exponent
$\alpha_{3D}$ of the 3D Ising model is identical to $\alpha_{2D}$, i.e. $\alpha_{3D}=0$. According to the scaling laws $d\nu=2-\alpha$ and
$\mu+\nu=2-\alpha$ [8], we have $\nu_{3D}=\frac{2}{3}$ and $\mu_{3D}=\frac{4}{3}$.

In conclusion, there is no question that Jacques H.H. Perk$^\prime$s comment [3] is definitely wrong due to misunderstanding my work.

This work was supported by the Sichuan Normal University and the "Thousand Talents Program" of Sichuan Province, China.

\vspace{2cm}

{\bf Reply to "Reply to Symmetry 2023, 15, 375"}

\vspace{0.5cm}
Note that
$\sum_{a=1}^{n}\sigma^z_{p+(a-1)m}(\prod_{i=p+(a-1)m+1}^{p+am-1}\sigma^x_i)\sigma^z_{p+am}
=\sum_{a=1}^{n}\sigma^z_{p+(a-1)m}{\sigma^\prime}^z_{p+am}\equiv\sum_{a=1}^{n}\sigma^z_{p+(a-1)m}\sigma^z_{p+am}$, which cannot lead to
$\sigma^z_{p+(a-1)m}{\sigma^\prime}^z_{p+am}=
\sigma^z_{p+(a-1)m}\sigma^z_{p+am}$. Obviously, such a person misunderstands the summation over $a$,
which implies the sum of spin configurations $(\pm 1)$ on $a$ [1,2,9]. Jacques H.H. Perk needs to carefully read
the Onsager$^\prime$s paper [2] and my work [1,9].


\end{document}